\documentclass[a4paper]{article}
\usepackage{RR}
\RRNo{6461}
\usepackage{hyperref}
\usepackage{mathptmx} 
\usepackage{graphicx}
\usepackage{algorithm}
\usepackage{algorithmic}
\usepackage{array}

\usepackage{parskip}

\usepackage{picinpar}

\RRdate{Juin 2006}
\RRauthor{
Pascal Barla
  \thanks[artis]{ARTIS GRAVIR/IMAG INRIA}%
  \and
Simon Breslav
\thanks[lee]{University of Michigan}%
\and
Lee Markosian
\thanksref{lee}
\and
Jo\"elle Thollot
\thanksref{artis}
}
\authorhead{Pascal Barla}
\RRtitle{Hachurage et pointillage par l'exemple}
\RRetitle{Interactive Hatching and Stippling by Example}
\titlehead{Interactive Hatching and Stippling by Example}
\RRresume{Ce rapport présente une méthode permettant à un artiste de
  dessiner interactivement un motif 2D de hachures ou de points puis de
  guider la synthèse d'un motif similaire. La synthèse s'appuie sur une
  phase d'analyse assistée par l'utilisateur dans laquelle le système
  extrait et organise des points ou des hachures (segments) selon des
  critères de regroupement 
  perceptuel. La synthèse est alors effectuée en combinant les
  propriétés (longueur, orientation, parallélisme, proximité) des
  éléments extraits par l'analyse.
}

\RRabstract{We describe a system that lets a designer interactively draw
 patterns of strokes in the picture plane, then guide the
 synthesis of similar patterns over new picture regions.
 Synthesis is based on an initial user-assisted analysis phase in
 which the system recognizes distinct types of strokes (hatching
 and stippling) and organizes them according to perceptual
 grouping criteria. The synthesized strokes are produced by
 combining properties (\eg, length, orientation, parallelism,
 proximity) of the stroke groups extracted from the input
 examples. We illustrate our technique with a drawing application
 that allows the control of attributes and scale-dependent reproduction of
 the synthesized patterns.}
\RRmotcle{Rendu expressif}
\RRkeyword{Expressive rendering, NPR}
\RRprojet{ARTIS}  
\RRtheme{\THCog} 
\URRhoneAlpes 

\newcommand{\ignorethis } [1] { }

\newcommand{\sectnum    } [1] {\ref{#1}}

\newcommand{\fignum     } [1] {\ref{#1}}

\newcommand{\sect       } [1] {Section~\sectnum{#1}}

\newcommand{\fig        } [1] {Figure~\fignum{#1}}

\newcommand{\etal       }     {\textit{et~al.}}

\newcommand{\eg         }     {{e.g.}}

\newcommand{\Ie         }     {{I.e.}}

\newcommand{\Reals      }     {{\textrm{I\kern-0.18em R}}}

\newcommand{\change     } [1] {\mbox{{\footnotesize $\Delta$} \kern-3pt}#1}

\begin{document}
\makeRR   

 \section{Introduction}
 \label{s:intro}

 \subsection{Motivation}
 \label{s:motivation}

 An important challenge facing researchers in non-photorealistic
 rendering (NPR) is to develop hands-on tools that give artists direct
 control over the stylized rendering applied to drawings or 3D
 scenes. An additional challenge is to augment direct control with a
 degree of \emph{automation}, to relieve the artist of the burden of
 stylizing every element of complex scenes. This is especially true
 for scenes that incorporate significant repetition within the
 stylized elements. While many methods have been developed to achieve
 such automation algorithmically outside of NPR (\eg, procedural
 textures), these kind of techniques are not appropriate for many NPR
 styles where the stylization, directly input by the artist, is not
 easily translated into an algorithmic representation. An important
 open problem in NPR research is thus to develop methods to analyze
 and synthesize artists' interactive input.

 In this work, we focus on the synthesis of stroke patterns that
 represent \emph{tone} and/or \emph{texture}. This particular class of
 drawing primitives have been investigated in the past (\eg,
 \cite{Salisbury:1994:IPI,Winkenbach:1994:CPI,Ostromoukhov:1999:DFE,Deussen:2000:FPA,Durand:2001:DSA}),
 but with the goal of accurately representing tone and/or texture
 coming from a photograph or a drawing. Instead, we orient our research
 towards the faithful reproduction of the expressiveness, or
 \emph{style}, of an example drawn by the user, and to this end
 analyze the most common stroke patterns found in illustration, comics
 or traditional animation: \emph{hatching} and \emph{stippling}
 patterns.

 \ignorethis{
 For example, an artist might paint a hatching pattern to represent a
 shadowed surface by carefully drawing a sequence of stroke layers
 that represent the tone of the depicted region. After doing all that
 work, it is natural for the artist to wish to replicate the pattern
 over new regions. In some cases, it might also be useful to generate
 a new version of the same pattern at a different resolution (for
 scale-dependent reproductions or levels of detail).  One way to
 achieve this is to simply copy and paste the stroke pattern
 repeatedly until the new region is covered. However, the resulting
 stroke pattern will generally exhibit obvious repetition, introducing
 a mechanical quality that was not present in the original pattern. To
 hide such repetition there is a need to introduce variation in the
 generated patterns.
}

 Our goal is thus to synthesize stroke patterns that ''look like'' an
 example pattern input by the artist, and since the only available
 evaluation method of such a process is visual inspection, we need to
 give some insights into the perceptual phenomena arising from the
 observation of a hatching or stippling pattern.  In the early 20th
 century, Gestalt psychologists came up with a theory of how the human
 visual system structures pictorial information. They showed that the
 visual system first extracts atomic elements (\eg, lines, points, and
 curves), and then structures them according to various perceptual
 grouping criteria like proximity, parallelism, continuation,
 symmetry, similarity of color, velocity, etc. This body of research
 has grown consequently under the name of \emph{perceptual
 organization} (see for example the proceedings of POCV, the IEEE
 Workshop on Perceptual Organization in Computer Vision). We believe
 it is of particular importance when studying artists' inputs.

 \subsection{Related work}
 \label{s:related}

  The idea of synthesizing textures, both for 2D images and 3D
  surfaces, has been extensively addressed in recent years (\eg~by
  Efros and Leung \cite{Efros:1999}, Turk
  \cite{Turk:2001:TSO}, and Wei and Levoy
  \cite{Wei:2001:TSO}). Note, however, that this body of research
  is concerned with painting and synthesizing textures that are
  represented as \emph{images}.  In contrast, we are concerned with
  direct painting and synthesis of stroke patterns represented in
  \emph{vector} form. \Ie, the stroke geometry is represented
  explicitly as connected vertices with attributes such as width and
  color. While this vector representation is typically less efficient
  to render, it has the important advantage that strokes can be
  controlled procedurally to adapt to changes in the depicted regions
  (strokes can vary in opacity, thickness and/or density to depict an
  underlying tone.)

  Stroke pattern synthesis systems have been studied in the past, for
  example to generate stipple drawings \cite{Deussen:2000:FPA},
  pen and ink representations \cite{Salisbury:1994:IPI,Winkenbach:1994:CPI},
  engravings \cite{Ostromoukhov:1999:DFE}, or for painterly rendering
  \cite{Hertzmann:1998:PRW}.
  However, they have relied primarily on generative rules, either
  chosen by the authors or borrowed from traditional drawing
  techniques. We are more interested in analysing reference patterns
  drawn by the user and synthesizing new ones with similar perceptual
  properties.

  Kalnins \etal~\cite{Kalnins:2002:WND} described an algorithm
  for synthesizing stroke ``offsets'' (deviations from an
  underlying smooth path) to generate new strokes with a similar
  appearance to those in a given example set. Hertzmann
  \etal~\cite{Hertzmann:2002:CA}, as well as Freeman 
  \etal~\cite{Freeman:2003:style} address a similar
  problem. Neither method reproduces the inter-relation of
  strokes within a pattern. Jodoin \etal~\cite{Jodoin:2002:HBE}
  focus on synthesizing hatching strokes, which is a relatively
  simple case in which strokes are arranged in a linear order
  along a path. The more general problem
  of reproducing organized patterns of strokes has remained an
  open problem.

 \subsection{Overview}
 \label{s:overview}

 In this paper, we present a new approach to analyze and synthesize
 hatching and stippling patterns in 1D and 2D. Our method relies on
 user-assisted analysis and synthesis techniques that can be governed
 by different behaviors. In every case, we maintain low-level
 perceptual properties between the reference and synthesized patterns
 and provide algorithms that execute at interactive rates to allow the
 user to intuitively guide the synthesis process.

 The rest of the paper is organized as follows. We describe the
 analysis phase in \sect{s:analysis}, and the synthesis algorithm and
 associated ``behaviors'' in \sect{s:synthesis}. We present results in
 \sect{s:results}, and conclude in \sect{s:discussion} with a
 discussion of our method and possible future directions.

  \section{Analysis}
  \label{s:analysis}

\begin{figure}[t]
   \begin{center}
     \begin{tabular}{c}
       \includegraphics[width=5cm]{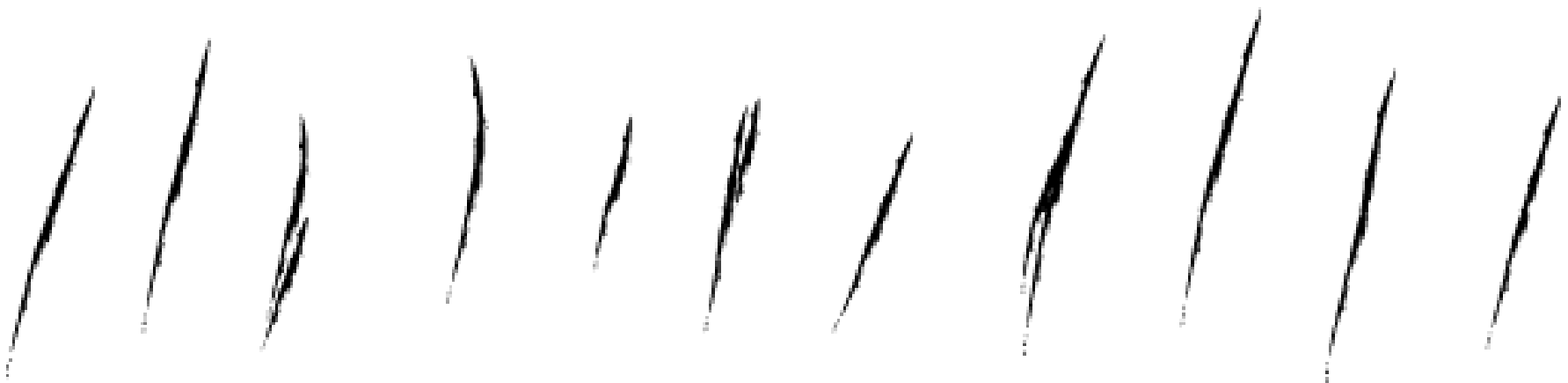} \\ \hspace{-0.4cm} \vspace{-0.2cm}
       \includegraphics[width=5cm]{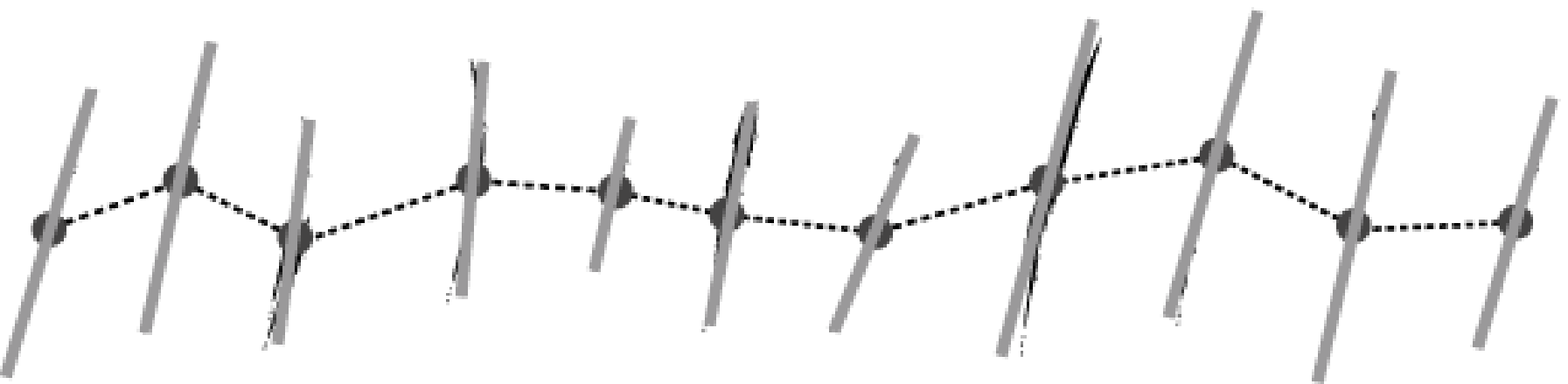}
     \end{tabular}
   \end{center}
   \caption{\label{f:anal_ex}A simple example of the analysis
     process in 1D: the strokes input by the user (top) are
     analyzed to extract elements (bottom, in light gray), 
     that are further organized along a path (dashed polyline).}
 \end{figure}

  We structure a stroke pattern according to perceptual organization
  principles: a pattern is a \emph{collection} of groups (hatching or
  stippling); a group is a \emph{distribution} of elements (points or
  lines); and an element is a \emph{cluster} of strokes.  For instance,
  the user can draw a pattern like the one in \fig{f:anal_ex}, which is
  composed of sketched line segments, sometimes with a single stroke,
  sometimes with multiple overlapping strokes; our system then clusters
  the strokes in line elements that hold specific properties; and finally
  structures the elements into a hatching group that holds its own
  properties.  We restrict our analysis to homogeneous groups with an
  approximate uniform distribution of their elements: hatching groups are
  made only of lines, stippling groups made only of points. This approach
  could be extended to more complex elements, using the clustering technique 
  of Barla \etal~\cite{BTS05a}.

  As a general rule of thumb, we consider that involving the user in the
  analysis gives him or her more control over the final result, at the
  same time removing complex ambiguities.  Thus, in our system, the user
  first specifies the high-level properties of the stroke pattern he is
  going to describe. He chooses a type of pattern (hatching or stippling);
  this determines the type of elements to be analyzed (lines for hatching,
  points for stippling). He then chooses a 1D or 2D reference frame within 
  which the elements will be placed. He finally sets the scale $\epsilon$ of 
  the elements, measured in pixels: intuitively, $\epsilon$ represents the
  maximum diameter of analysed points, and the maximum thickness of
  analyzed lines. 

  Once these parameters are set, the user draws strokes as polyline
  gestures. Depending on the group type, points or lines at the scale
  $\epsilon$ are extracted and structured: Then statistics about perceptual
  properties of the strokes are computed. This whole processus has an instant 
  feedback, so that the user can vary $\epsilon$ and observe changes made to the
  analysis in real-time. We first describe how elements are extracted given a chosen
  $\epsilon$ and their analyzed properties; then we describe how those
  elements are structured into a group, and how perceptual measures that
  characterize this group are extracted.

  \subsection{Element analysis}
  \label{s:element}

\begin{figure}[ht]
   \begin{center}
     \begin{tabular}{cc}
       \includegraphics[height=0.8cm]{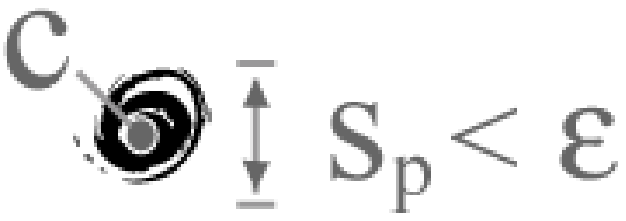} &\vspace{0.5cm}
       \includegraphics[height=0.8cm]{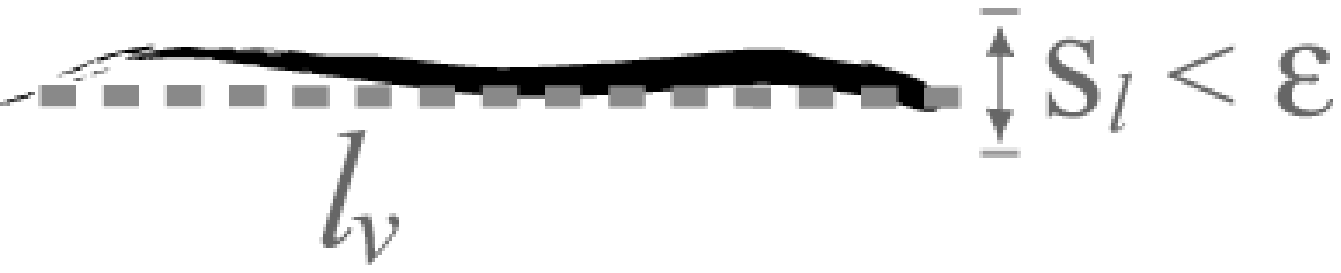} \\ 
       \includegraphics[height=0.8cm]{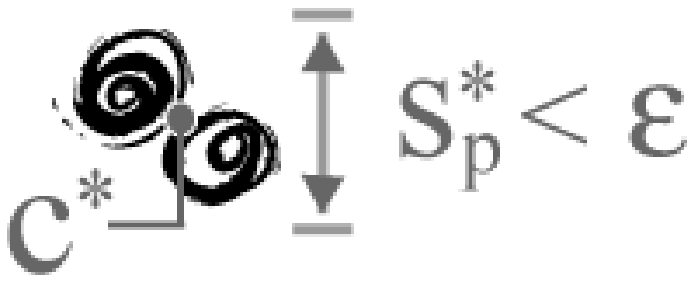} &
       \includegraphics[height=0.8cm]{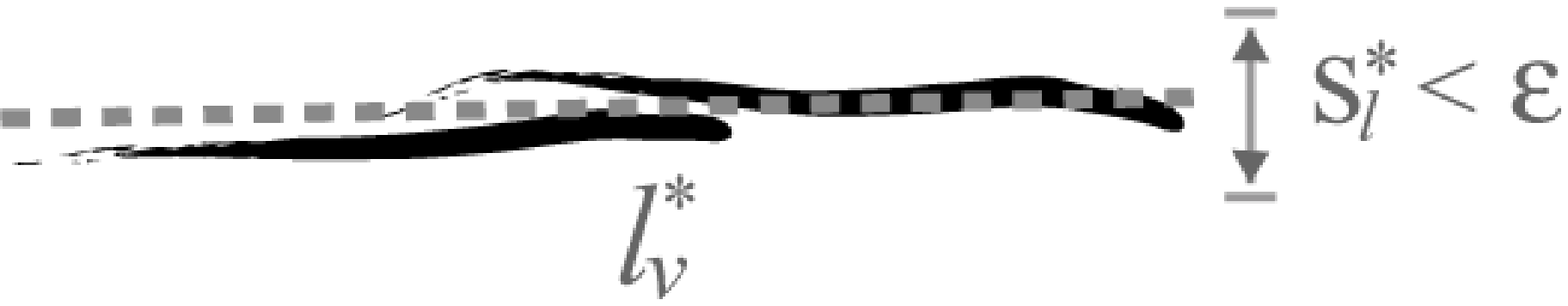} 
     \end{tabular}
   \end{center}
   \caption{ Top left: A stroke is fit to a point. Bottom left: A pair of points is clustered into 
     a new point. Top right: A stroke is fit to a line. Bottom right: A pair of lines is 
     clustered into a new line.}
   \label{f:element_analysis}
 \end{figure}

  The purpose of element analysis is to cluster a set of strokes drawn by
  the user into points or lines, depending on the chosen element type.  To
  this end, we use a greedy algorithm that processes strokes in the
  drawing order, and tries to cluster them until no more clustering can be
  done. We first fit each input stroke to an element (point or line) at
  the scale $\epsilon$.  Strokes that cannot be fit to an element are
  flaged \emph{invalid} and will be ignored in the remaining steps of the
  analysis. Then, valid pairs of elements that can be perceived as a
  single element are clustered iteratively. The fitting and clustering of
  points and lines is illustrated in \fig{f:element_analysis}.

  For points, the fitting is performed by
  computing the center of gravity $c$ of a stroke $S$ and measuring its
  spread $s_p=2\max_{p \in S} |p - c|$. If $s_p>\epsilon$, then the stroke
  is flaged \emph{invalid} because the circle of center $c$ and diameter
  $s_p$ do not encloses $S$. The clustering of two points
  is made by computing the center of gravity $c^*$
  of the points and measuring its spread $s_p^*$. Similarly, if
  $s_p^*>\epsilon$, then the points cannot be clustered. This allows the
  system to recognize any cluster of short strokes relative to the scale
  $\epsilon$, like point clusters, small circled shapes, crosses,
  etc. (See \sect{s:results}.)

  For lines, the fitting is performed by computing the virtual line $l_v$
  of a stroke $S$ and measuring its spread $s_l=2\max(d_H(S,l_v),
  d_H(l_v,S))$ where $d_H(X,Y)=\max_{x \in X} (\min_{y \in Y} |x - y|)$ is
  the Hausdorff distance between two sets of points.  The virtual line can be computed by least-square
  fitting, but in practice we found that using the endpoint line is enough
  and faster. Then, if $s_l>\epsilon$, the stroke is flaged
  \emph{invalid} because the line segment of axis $l_v$ and thickness
  $s_l$ do not enclose $S$. The clustering of two lines
  is made by computing the virtual line $l_v^*$ of
  the lines and measuring its spread $s_l^*$. The virtual line can be
  computed by least-square fitting on the whole set of points; but we
  preferred to apply least-square fitting only on the two endpoints of each clustered
  line for efficiency reasons.  Then, if $s_l^*>\epsilon$, the lines
  cannot be clustered.  This allows the system to recognize any set of
  strokes that resembles a line segment at the scale $\epsilon$. Examples
  including sketched lines, overlapping lines, and dashed lines are shown
  in \sect{s:results}.

  Once points or lines have been extracted, we can compute their
  properties: extent, position and orientation. The extent property represents 
  the dimensions of the element: point size or line length and width.  
  For point size, we use the spread of the element. For lines, we use the length of the
  virtual line and its spread (for width).  Orientation represents the
  angle between a line and the reference frame main direction (the main
  axis for 1D frames, the X-axis for the cartesian frame). It is always
  ignored for points.  We add a special position property for 1D reference
  frames: since they are synthesized in 2D (in the picture plane), 1D
  patterns have a remaining degree of freedom that is represented by the
  position of elements perpendicular to the main axis.  For all these
  properties, we compute statistics (a mean and a standard deviation) and
  boundary values (a min and a max); We also store the gesture input by the 
  user and will refer to it as the \emph{shape} of the element in the rest of the paper.

  \subsection{Group analysis}
  \label{s:group}

  A group is considered to be an approximately uniform distribution of
  elements within a reference frame. This means that while analyzing a
  reference group, we are not interested in the exact distribution of its
  elements: we consider a reference group as a small sample of a bigger,
  approximately uniform distribution of the same elements. Consequently,
  we first need to extract a local structure that describes the
  neighborhood of each element; this local structure will then be
  reproduced more or less uniformly throughout the pattern during
  synthesis.

  To this end, we begin with the computation of a graph that
  structures the elements locally: in 1D, we build a chain that
  orders strokes along the main axis; whereas in 2D, we compute a
  Delaunay triangulation. We only keep the edges that: (a)
  connect two \emph{valid} elements and (b) connect an element to
  its \emph{nearest neighbor}. \ignorethis{(see
  \fig{f:synth-graph})} We chose this because the synthesis
  algorithm (described in \sect{s:synthesis}) converges only when 
  considering nearest neighbor edges. However, this decision is also
  justified from a perceptual point of view: basing our analysis
  on nearest neighbors emphasizes the proximity property of
  element pairs, which is known to be a fundamental perceptual
  organization criterion.

  For each edge of the resulting graph, we extract the following
  perceptual properties, taking inspiration from Etemadi
  \etal~\cite{bb3357}: proximity for points and lines; parallelism,
  overlapping and separation for lines only. 
  
  \begin{window}[0,r,{\includegraphics[width=0.16\linewidth]{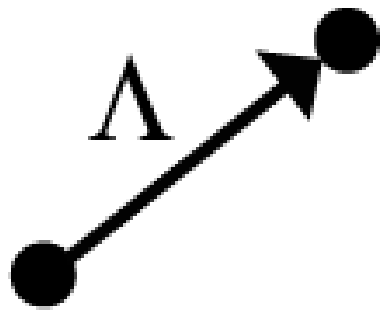}},{}] 
  Proximity is simply taken to be the euclidean distance between the centers of the
  two elements in pixels. We not only compute this measure for points, but also for lines 
  in order to initialize our synthesis algorithm (see \sect{s:synthesis}.)
  \end{window}
  Let's note $\Delta$ the vector from one point to the other,
  then we have 
  \begin{equation}
    prox=||\Lambda|| 
  \end{equation}
  with $prox \in [0,+\infty)$.

  \begin{window}[0,r,{\includegraphics[width=0.28\linewidth]{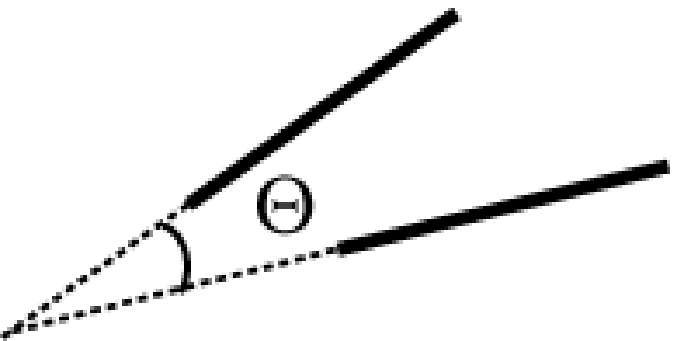}},{}] 
  To compute parallelism, we first find the accute angle made between the two lines.
  Since there is no apriori order on the line pair, we take the absolute value of 
  this accute angle and normalize it between $0$ and $1$. 
  \end{window}
  Let's note $\Theta$ the accute angle, then we compute parallelism using
  \begin{equation}
    par=|\frac{2\Theta}{\pi}|
  \end{equation}
  with $par \in [0,1]$.

  \begin{window}[0,r,{\includegraphics[width=0.294\linewidth]{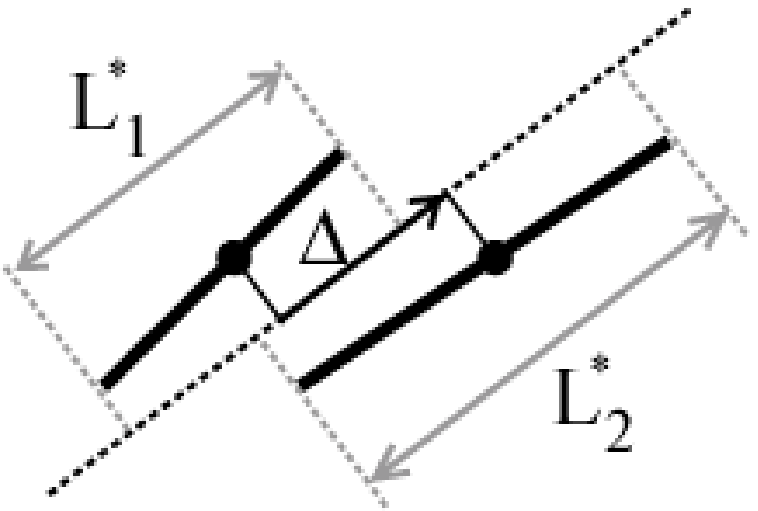}},{}] 
  Like Etemaldi \etal~\cite{bb3357}, we define overlapping relative
  to the bissector of the considered line pair. But we modify slightly their
  measure to meet our needs: We project the center of each line on the bissector
  and use them to define an overlapping vector $\Delta$. 
  \end{window}
  Overlapping is computed using the following formula:
  \begin{equation}
    ov=\frac{2||\Delta||}{L_1^*+L_2^*}
  \end{equation}
  where $L_1^*$ and $L_2^*$   are the lengths of the lines projected on the bissector.
  Note that with this definition, $ov=0$ means a perfect overlapping.

  \begin{window}[0,r,{\includegraphics[width=0.29\linewidth]{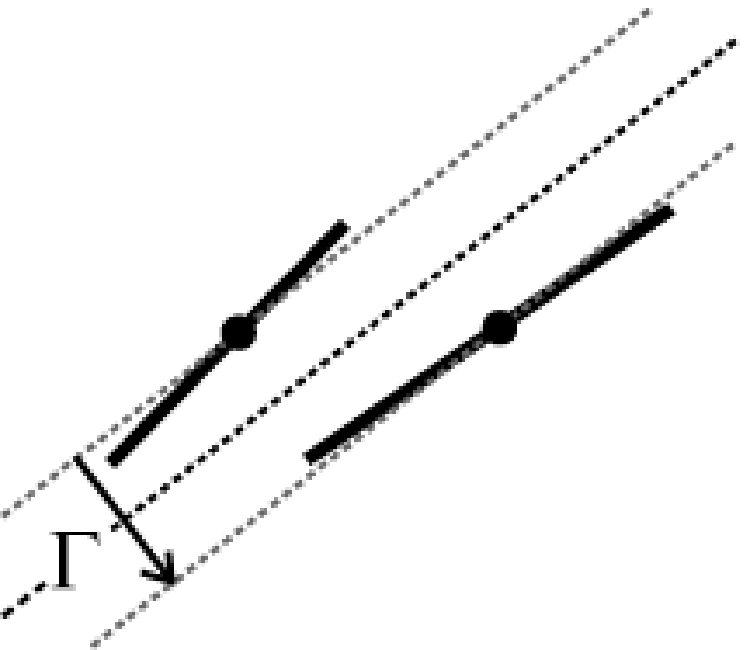}},{}] 
  Finally, separation represents the distance between two lines, this time in the direction 
  perpendicular to their bissector. We project the center of each line on a line perpendicular to the 
  bissector and use them to define a separation vector $\Gamma$.
  \end{window}
  Separation is then computed with the following formula:
  \begin{equation}
    sep=||\Gamma||
  \end{equation}
  with $sep \in [0,+\infty)$.

  We compute statistics (a mean and a  standard deviation) and bounds (a min and a max) for each of these properties.

  \section{Synthesis}
  \label{s:synthesis}

  The purpose of the synthesis process is to create a new stroke pattern
  that has the same properties (for elements and groups) as the reference
  pattern. We first describe a general algorithm that is able to create a
  new pattern meeting this objective; then we show how to customize it
  through the use of synthesis \emph{behaviors}.

  \subsection{Algorithm}
  \label{s:algorithm}

  Our synthesis algorithm can be summarized as follows:
  \begin{enumerate}
    \item Build a graph where the edge lengths follow the proximity 
      statistics;
    \item Synthesize an element at each graph node using element properties;
    \item Correct elements position and orientation using element pair properties.
  \end{enumerate}

  The first step is achieved using Lloyd relaxation \cite{lloyd1}. This
  technique starts with a random distribution of points in 1D or 2D. It
  then computes the Voronoi diagram of the set of points, and moves each
  point to the center of its Voronoi region.  When applied iteratively,
  the algorithm converges to an even distribution of points.  Deussen
  \etal~\cite{Deussen:2000:FPA} observed that the variance of nearest
  neighbor edge length decreases with each iteration. We use this to get a
  variance (in nearest neighbor edge length) that approximately matches
  that of the reference pattern.

\begin{figure}[ht]
   \begin{center} 
     \begin{tabular}{cc}\hspace{-0.2cm}
       \includegraphics[height=2.8cm]{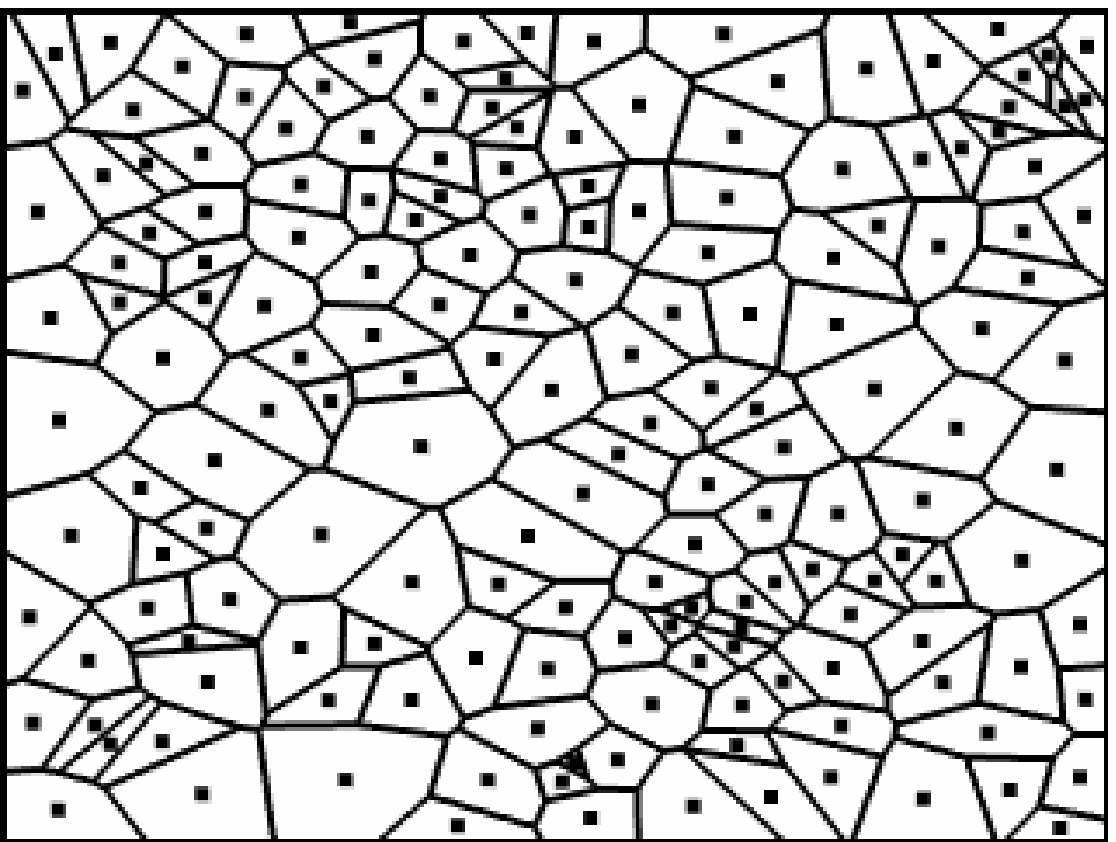} & 
       \includegraphics[height=2.8cm]{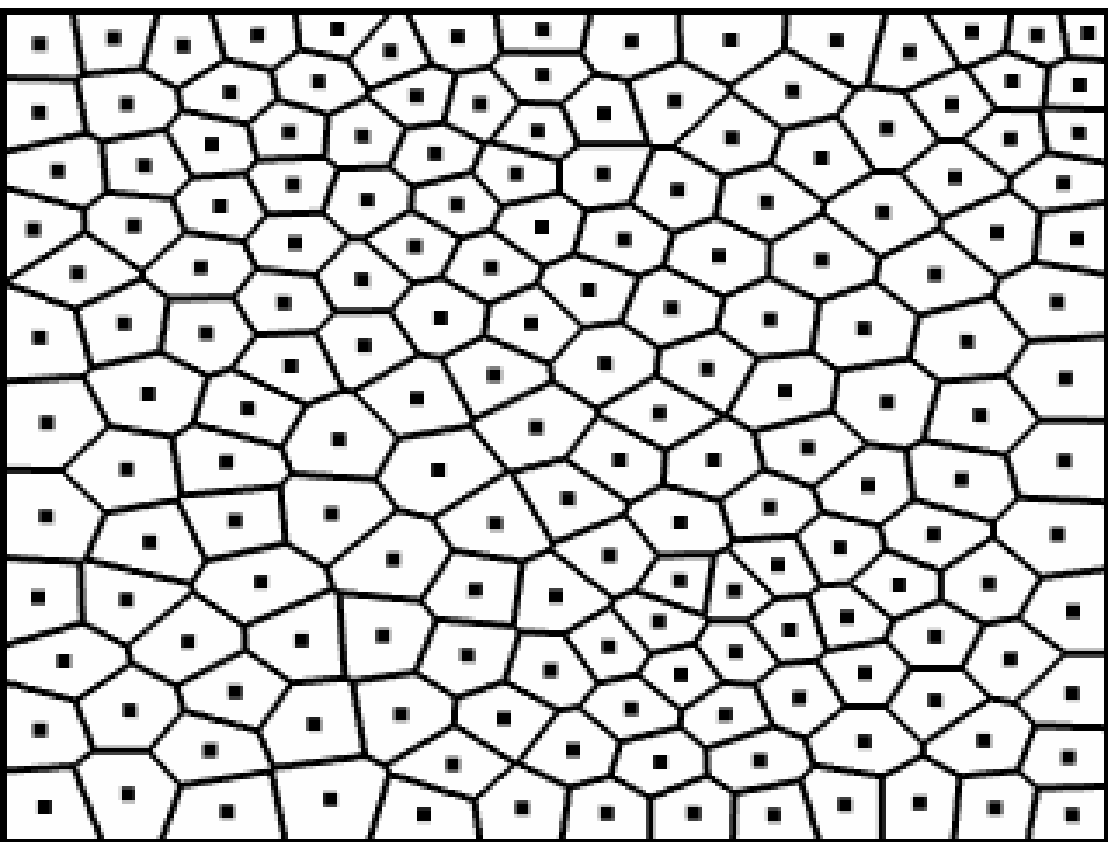} \\
       (a) & (b)
     \end{tabular}
   \end{center}
   \caption{\label{f:lloyd}(a) An input random distribution and
     its Voronoi diagram. (b) The result after iteratively
     applying Lloyd's method until a desired variance-to-mean
     ratio in edge length is obtained. }
 \end{figure}

  Consider the mean $\mu^*$, standard deviation $\sigma^*$, and the ratio
 $r^*=\sigma^*/\mu^*$~of a given property in our reference pattern. We
  start with a random point set by distributing $N =
  N_{ref}\mathcal{A}/\mathcal{A}_{ref}$ points, where $N_{ref}$ is the
  number of elements in the reference pattern, and $\mathcal{A}_{ref}$ and
  $\mathcal{A}$ are the area of the reference and target patterns,
  respectively. We then apply the Lloyd technique, computing $\mu$,
  $\sigma$ and $r=\sigma/\mu$ of the current distribution at each step
  until $r<r^*$.  Note that $\mu$ will have changed throughout the set of
  iterations. Thus, in order to have $\mu = \mu^*$, we finally rescale the
  distribution by $\mu^*/\mu$. An example of Lloyd's method is shown in
  \fig{f:lloyd}.

  In the second step, for each node of the graph we first pick a reference
  element $E$.  Then we choose a set of element properties and compute a
  position, orientation and scale for $E$. There are many different
  approaches to choose element properties; the ones we implemented are
  detailed in the next section, and for now we only present the general
  algorithm.

  We first position the center of $E$ at its corresponding node
  location. In the case of a 1D reference frame, we also move $E$
  perpendicularly to the main axis using the relative position property.
  Then, $E$ is scaled using the extent property; however, we impose a
  constraint on scaling for each type of element.  In order for points to
  remain points, we ensure that their size is smaller than $\epsilon$; and
  similarly for lines, we ensure that their width is no more than
  $\epsilon$.  Finally, $E$ is rotated based on orientation. For a 1D
  reference frame, we rotate $E$ so that the angle with the local X-axis
  matches the orientation property. For a 2D reference frame, we use the
  angle with the global X-axis instead.

  Finally, in the third step, for each node of the graph, we compute a corrected set of
  parameters that takes into account the perceptual properties of nearest
  neighbor pairs extracted from the reference pattern during analysis. We use 
  a greedy algorithm where each node is corrected toward its nearest neighbor in turn.
  In order to get a consistent correction, we add two procedures to this algorithm: first, 
  the nodes are sorted according to the proximity with their nearest neighbor in a preprocess, 
  so that the perceptually closest elements are corrected in priority; second, when a node is corrected, 
  we discard both nodes of its edge from upcoming corrections, in order to ensure that the current 
  correction stays valid throughout the algorithm.

  We now describe how an element is corrected based on perceptual measures.
  In a way similar to what we did for element properties, we choose a set of
  perceptual properties for element pairs. The details of how we perform
  this choice are explained in the next section. Note that the correction
  is not directly applied to the initial set of parameters: the user can
  control through linear interpolation the amount of correction he or she
  wishes to apply.

   \begin{figure}[ht]
   \begin{center}
     \begin{tabular}{cc}\hspace{0.2cm}
       \includegraphics[height=2cm]{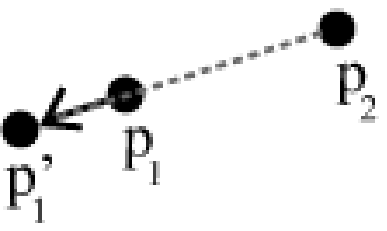} & \hspace{0.5cm}
       \includegraphics[height=2cm]{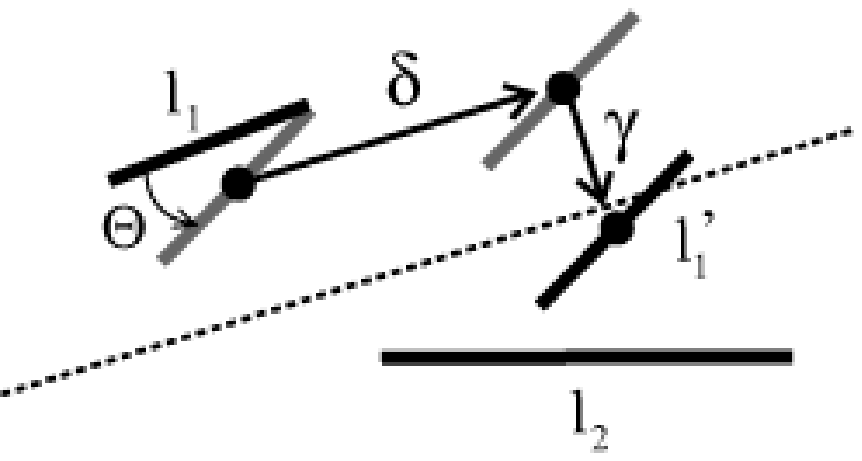} \\ \hspace{0.2cm} 
       (a) & \hspace{0.5cm} (b)
     \end{tabular}
   \end{center}
   \caption{ (a) A point is corrected by displacing it along the direction
   to its nearest neighbor to match a proximity measure. (b) A line is first 
   rotated to match parallelism, then displaced along and perpendicularly to 
   the bissector direction to match overlapping and separation.}
   \label{f:correct}
 \end{figure}

  For points, the only parameter to correct is position: we simply move
  the selected point along the line through its nearest neighbor to match
  the desired proximity (see \fig{f:correct}(a).) Let's consider $prox_1$ and $prox_2$, the current 
  and desired proximity values respectively. Then the correction applied to the
  position of current point is given by the following translation vector:
  \begin{equation}
    \lambda_{prox_1 \rightarrow prox_2} =  \frac{\Lambda_1}{||\Lambda_1||}.(prox_2-prox_1)
  \end{equation}

  If we are dealing with lines, we first correct the orientation of the current
  element based on the parallelism property we want to enforce; then we correct
  its position using overlapping and separation (see \fig{f:correct}(a).) The reason
  why we  first correct the orientation is that the overlapping property is highly
  dependent on the parallelism of lines.

  Let's consider $par_1$ and $par_2$, the current and desired parallelism values
  respectivelly. Then the correction applied to the orientation of current point 
  is given by the following angle:
  \begin{equation}
    \theta_{par_1 \rightarrow par_2} = sign(\Theta_1).(par_2-par_1).\frac{\pi}{2}
  \end{equation}

  Finally, we correct lines using a combination of overlapping and separation.
  For two overlapping values $ov_1$ and $ov_2$ for the current and target
  position, we translate the current line along the bissector line using the
  following vector:
  \begin{equation}
  \delta_{ov_1\rightarrow ov_2} = \frac{\Delta_1}{||\Delta_1||}.(ov_2-ov_1).\frac{(L_1^*+L_2^*)}{2}
  \end{equation}
  Then we translate it perpendicularly to the bissector using:
  \begin{equation}
  \gamma_{sep_1\rightarrow sep_2} = \frac{-\Gamma_1}{||\Gamma_1||}.(sep_2-sep_1)
  \end{equation}
  Note that the last two operations do not change the parallelism property.

  \subsection{Behaviors}
  \label{s:behaviors}

  We now present the synthesis \emph{behaviors} that are responsible for
  assigning a value for each property. We developed several behaviors
  because we believe that the ability to synthesize patterns that are more or
  less close to the reference pattern is a desirable feature: it lets us
  balance \emph{fidelity} and \emph{variation} relative to the reference
  pattern.

  We thus implemented three behaviors: \emph{sampling}, \emph{copying} and
  \emph{cloning}, that range from close to the statistical distribution to
  close to the reference data. In the same spirit, we let the user choose
  the amount of correction that is applied. The correction
  results are displayed interactively.  We now turn to the description of
  the three behaviors.

  \paragraph{Sampling}
  This behavior produces patterns whose properties exhibit the same
  statistics as those of the reference pattern. For each property, we
  compute the mean and standard deviation of values in the input pattern
  to derive a Gaussian distribution function, then sample its inverse
  cumulative function to yield values that follow the distribution of the
  reference pattern. If the sampled value lies outside the range of values
  in the reference pattern, the sampling is repeated until a value in the
  original range of values is produced. The reference element whose shape
  is to be copied is then randomly chosen.

  \paragraph{Copying}
  Moving toward increased fidelity to the reference pattern, this behavior
  assigns each property independently by copying values from randomly
  chosen elements in the reference pattern.  For pairs, the reference pair
  with most similar value is found, and the synthesized pair is altered to
  match the reference pair. As an example, consider the proximity property
  of element pairs. If a nearest-neighbor pair of synthesized elements is
  separated by $n$ pixels, we first find the reference pair whose
  proximity $m$ is closest to $n$. We then correct the position of the
  chosen synthesized element to achieve a proximity of $m$ pixels. 
  For element shape, we first pick the reference pair with the most
  similiar proximity value and copy one of its element randomly.

  \paragraph{Cloning}
  The cloning behavior synthesizes patterns that most closely follow the
  reference pattern.  It is a modification of the copy behavior where all
  the properties are taken from the same source. Given a synthesized
  element, we randomly choose a reference element and copy \emph{all} its
  properties to the synthesized element. Pairs are handled similarly, but
  the choice is not random: during correction, in order to stay close to
  the statistical distribution produced by Lloyd's method, we first find
  the reference pair with the most similar proximity property, then adjust
  parameters of the chosen synthesized element to yield the same pair-wise
  properties. The element shape is chosen like with the copying behavior.

  \section{Results}
  \label{s:results}

\begin{figure}[h]
   \begin{center}
     \begin{tabular}{lc}
       \raisebox{1.5ex}{(a)} & \includegraphics[width=8.5cm]{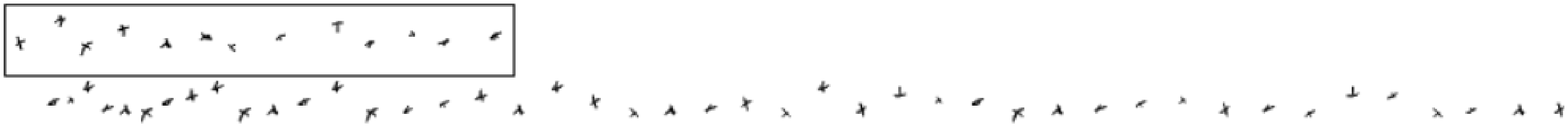} \vspace{0.3cm} \\ 
       \raisebox{1.5ex}{(b)} & \includegraphics[width=8.5cm]{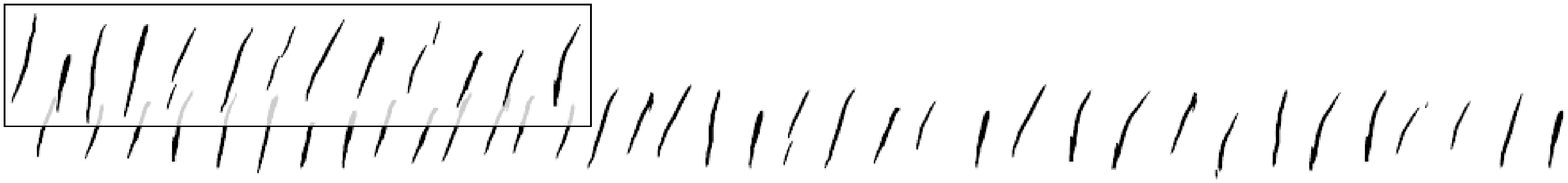} \vspace{0.2cm}\\ 
       \raisebox{8ex}{(c)} & \includegraphics[width=8.5cm]{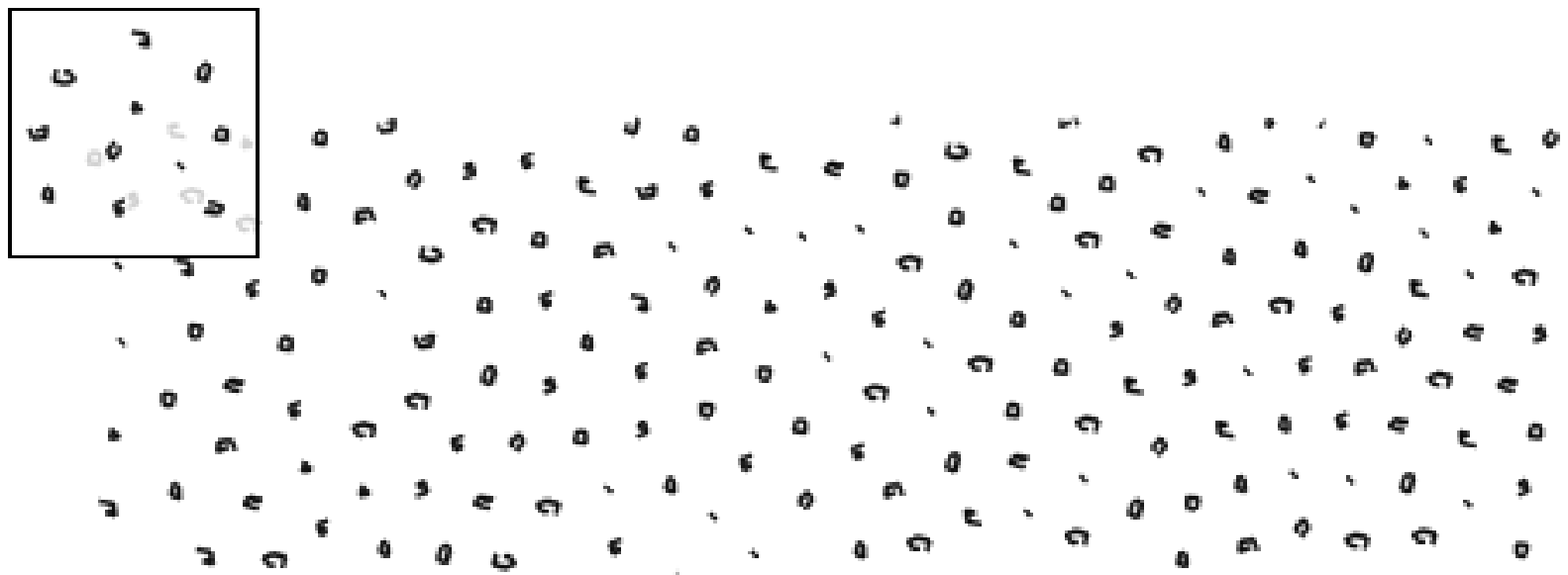} \vspace{0.1cm}\\ 
       \raisebox{9ex}{(d)} & \includegraphics[width=8.5cm]{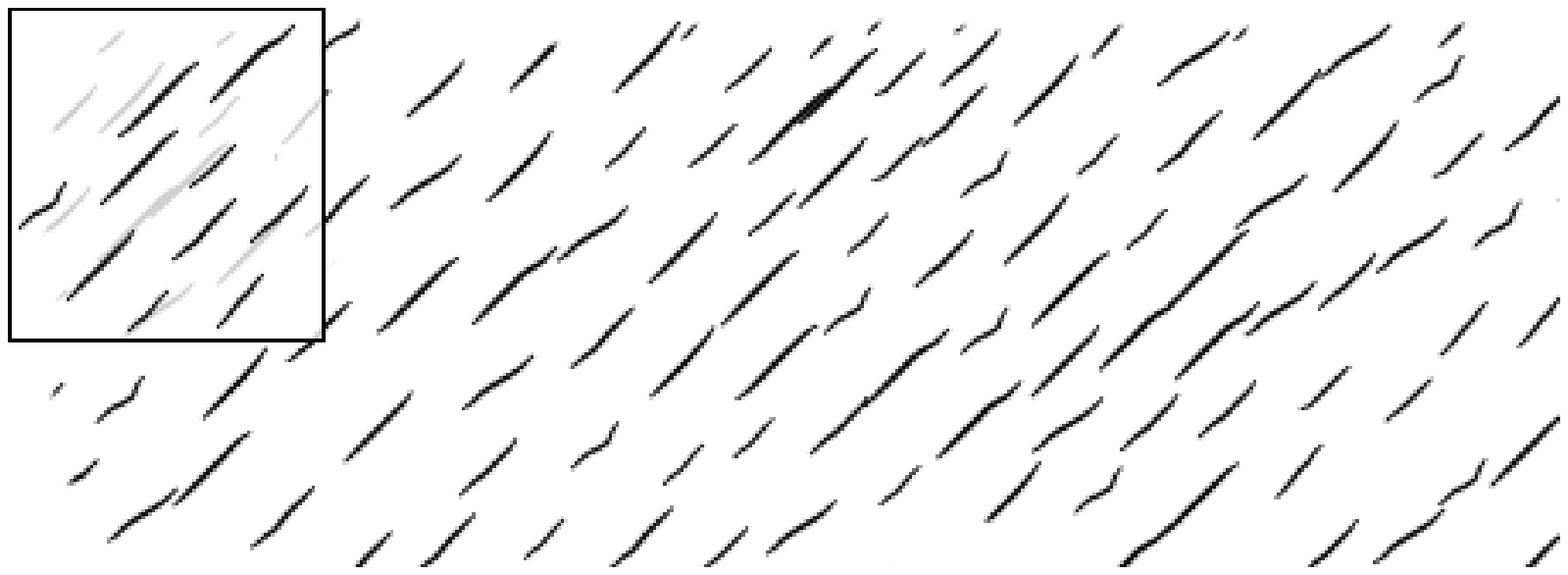} \vspace{0.1cm}\\
       \raisebox{1.5ex}{(e)} & \includegraphics[width=8.5cm]{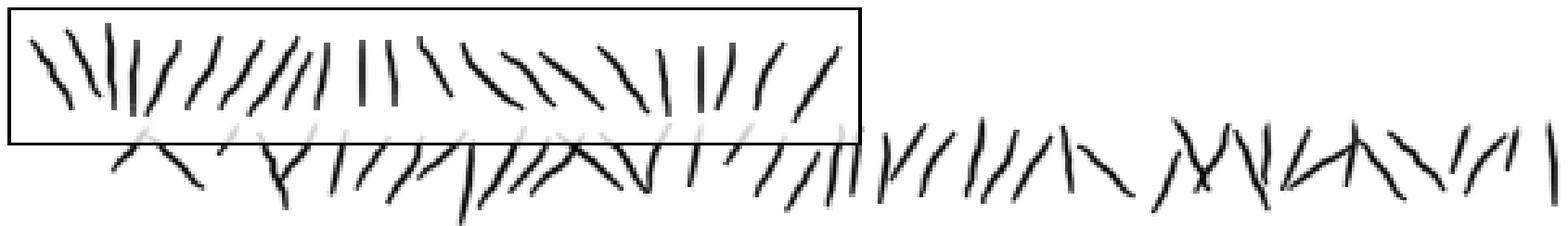} \vspace{0.2cm}\\
       \raisebox{6.5ex}{(f)} & \includegraphics[width=8.5cm]{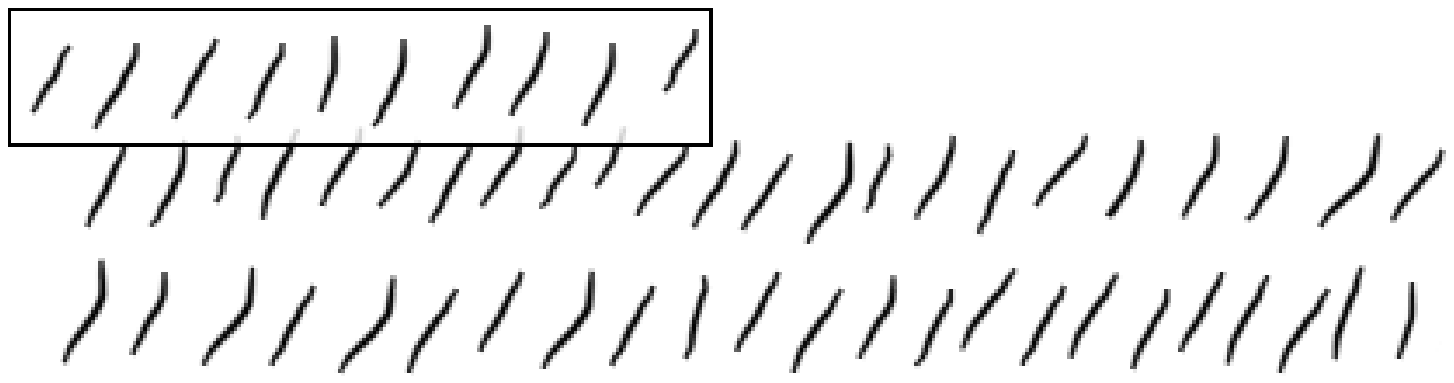} \vspace{0.2cm}\\
       \raisebox{7.5ex}{(g)} & \includegraphics[width=8.5cm]{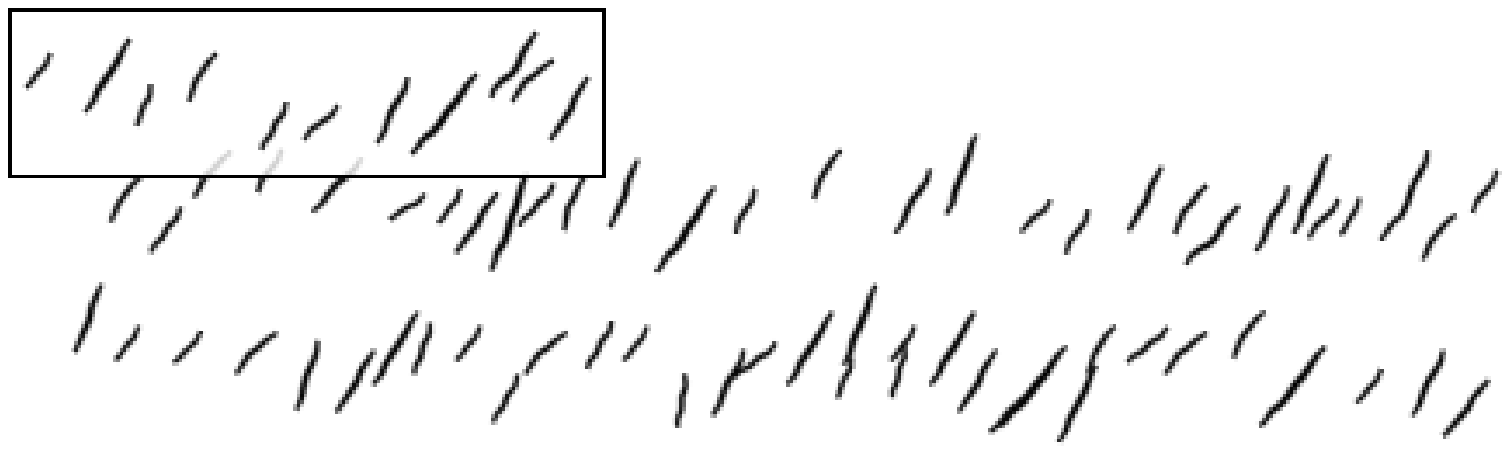}
     \end{tabular}
   \end{center}
   \caption{ (a)-(d) Stippling and hatching results in 1D and 2D with their reference 
   pattern (small boxes); (e) A synthesis failure example; (f) For uniform patterns, 
   sampling (top) introduces more variation than cloning (bottom); (g) For less uniform 
   patterns, sampling (top) is more incoherent than cloning (bottom).}
   \label{f:results}
 \end{figure}

  \fig{f:results}(a)-(d) shows reference and synthesized hatching and
  stippling groups in 1D and 2D. For these examples, we used the copying
  behavior and chose the correction amount manually. Note that complex elements extracted
  in the analysis phase are reproduced during synthesis: crosses and small
  circles for stippling, sketched strokes and multiple overlapping lines
  for hatching. The relations among nearest-neighbor synthesized elements
  are replicated from corresponding reference elements. \fig{f:results}(e)
  shows a limitation of our method: the synthesis fails to reproduce
  recognizable stroke sequences seen in the input pattern. This is due to
  the small neighborhood size (only nearest neighbor) used in synthesis.
  The computation times are interactive, up to a couple of seconds for 
  the most complex patterns we synthesized.

  We compare our different behaviors in \fig{f:results}(f)-(g), using two
  reference 1D hatching patterns. With a quasi-uniform pattern (similar
  orientation, spacing, etc.), the sampling behavior has the advantage of
  synthesizing patterns with more variation, creating strokes in positions
  and orientations that were not present in the reference pattern, whereas
  the cloning behavior reproduces strokes and nearest neightbor relations
  found in the reference pattern.  With a more irregular reference
  pattern, the sampling behavior produces patterns that lack the broader
  coherence of the example pattern, while the cloning behavior synthesizes
  patterns with more fidelity. Although not shown here, the copying
  behavior produces intermediate results, providing a trade-off between
  fidelity and variation.

\begin{figure*}[ht]
   \begin{center}
     \begin{tabular}{cc} \hspace{-1cm}
     \includegraphics[height=8cm]{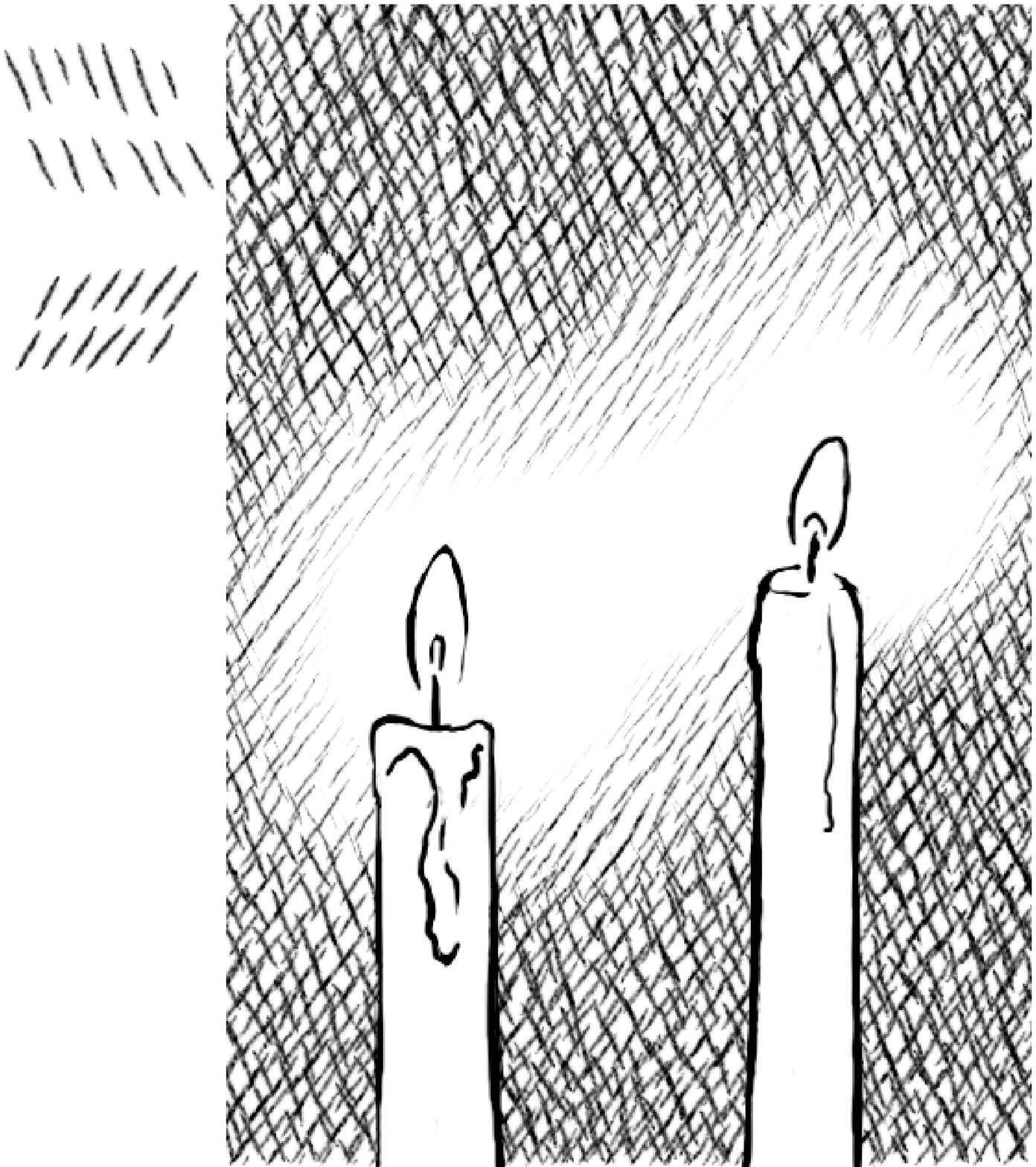} & 
     \includegraphics[height=8cm]{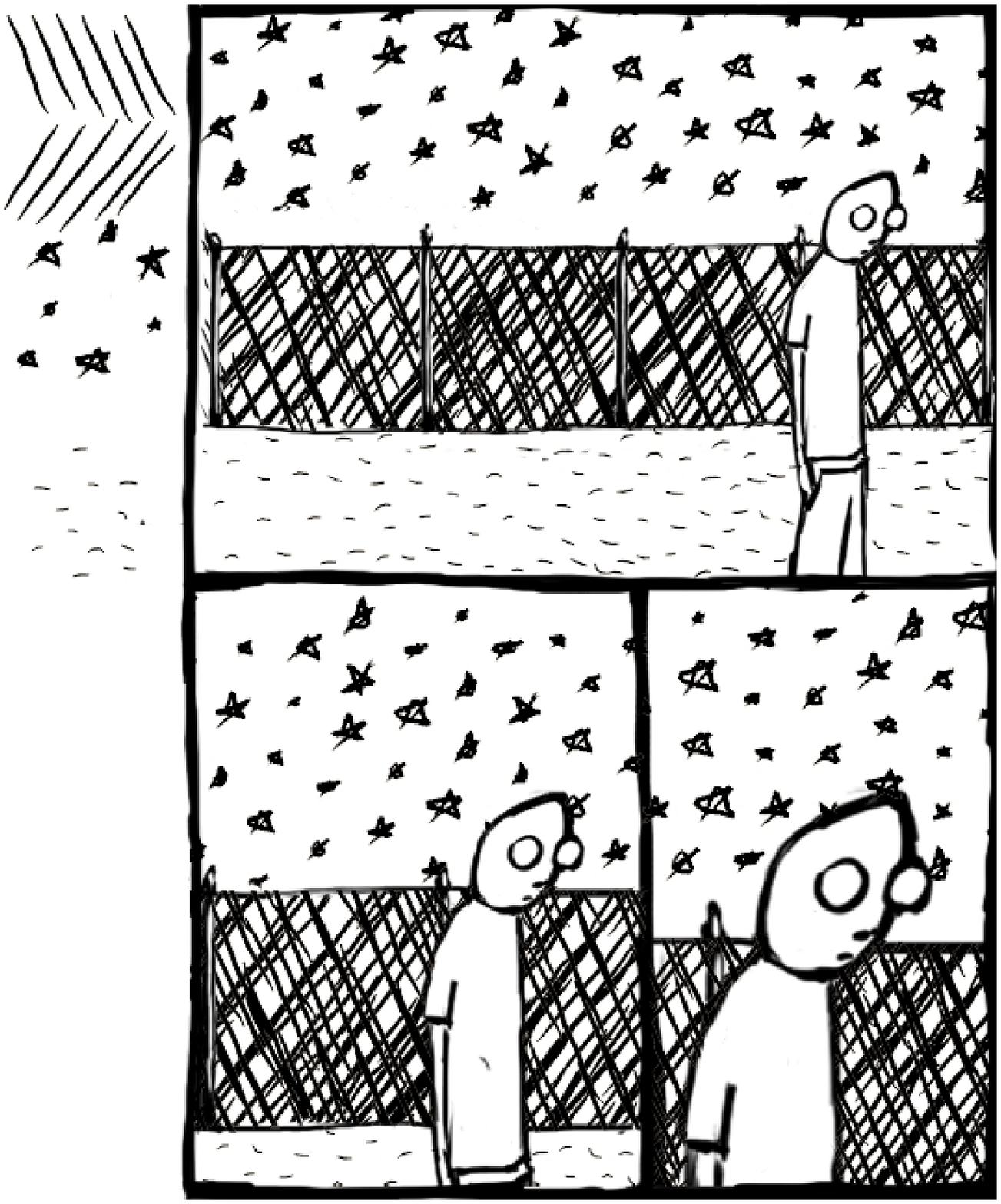}
     \end{tabular}
   \end{center}
   \caption{Two examples from our drawing application. Left: the thickness of
   the lines is controled by a user-defined mask; Right: levels-of-detail of
   a drawing (bottom) are automatically synthesized from an initial resolution (top).}
   \label{f:applications}
 \end{figure*}

  To illustrate our synthesis technique, we developed a 2D drawing
  application that lets the user draw example hatching or stippling
  patterns, then guide the synthesis of similar patterns over selected
  image regions. The system can vary stroke attributes such as color,
  thickness and opacity according to colors or tones in a provided
  background image (see \fig{f:applications}, left). This lets us create colored
  strokes that represent shadows, highlights and intermediate tones. The
  final illustration is composed of the synthesized patterns, optionally
  composited on top of the background image. The system can also
  synthesize multiple versions of the same pattern at different
  resolutions, supporting the scale-dependent reproductions of the output
  image, for simple levels-of-detail or for printing purpose (see \fig{f:applications}, right).

  \section{Discussion and future work}
  \label{s:discussion}

  The main limitation of our approach is that we only consider nearest
  neighbor relations. To our knowledge, the first and only attempt to deal
  with larger neighborhoods is the work of Jodoin
  \etal~\cite{Jodoin:2002:HBE}. However, they only consider parametric
  elements of the same dimension, organized along a 1D path with no
  perceptual analysis (though they acknowledge the importance of
  extracting low-level perceptual properties). On the other hand, our
  approach is a step in another direction: we consider more general
  elements thanks to our element analysis, and introduce low-level
  perceptual properties in our group analysis. We also generalize the
  synthesis to 2D patterns. These two approaches are not incompatible,
  however. We look forward to combining advantages of both methods to
  develop a more general solution to the example-based stroke pattern
  synthesis problem.

  We believe that extending our method to bigger neighborhoods, exploiting
  perceptual properties for neighborhood comparisons, is the logical next
  step towards this solution. It is interesting to compare our method to
  work done on texture synthesis \emph{on surfaces} (\eg,
  Turk~\cite{Turk:2001:TSO} and Wei and Levoy~\cite{Wei:2001:TSO}). In
  those systems and ours, the first step is to build a lattice that models
  the correspondence between the input (2D texture or reference stroke
  pattern) and the output (a 2D surface or target distribution of
  elements). The second step initializes nodes of the lattice with values
  that follow the statistics of the reference pattern or texture. In the
  final step, node values are modified based on neighborhood comparisons
  of various sizes, either at random \cite{Wei:2001:TSO} or in a
  predefined order (line sweeping in \cite{Turk:2001:TSO}, nearest
  neighbors in our work). This observation opens promising avenues for
  building on the existing body of texture synthesis techniques.

  For our clustering algorithm, we relied on the drawing order, but we
  might investigate other orderings, for instance based on proximity. This
  would even be mandatory in cases where the reference pattern is
  extracted from an image and the drawing order is not known.  We also
  considered only points and lines and the perceptual relations among
  them. However, other primitives like arcs or more complex curves have
  been studied from a perceptual organization point of view, and we plan
  to incorporate them in our system in future work. Other perceptual
  criteria like symmetry or closeness might then be of great value for
  those patterns.  Finally, even if we believe that a user-assisted
  analysis is the most valuable approach, one might consider automating
  the process for specific applications (\eg~for capturing the style of an
  existing drawing). This implies determining the group type, reference
  frame and $\epsilon$ automatically.

\section{Conclusions}
\label{s:conclusions}
 
  We presented a new approach to stroke synthesis by example, for two
  particular classes of patterns: hatching and stippling (in 1D and 2D).
  Our method is fast and easy to implement. Its interactive response and
  its different synthesis behaviors let the user guide the synthesis
  process.  The resulting synthesized patterns are perceptually similar to
  the reference ones, but also add a degree of variation. This lets us use
  our tool in a drawing application, with features such as stroke attribute 
  control for efficient image depiction, and scale-dependent synthesis for
  levels-of-detail.


\bibliographystyle{alpha}
\bibliography{npr}

\newcommand{\etalchar}[1]{$^{#1}$}
\begin{thebibliography}{JEGPO02}

\bibitem[BTS05]{BTS05a}
Pascal Barla, Jo{\"e}lle Thollot, and Fran\c{c}ois Sillion.
\newblock Geometric clustering for line drawing simplification.
\newblock In {\em Proceedings of the Eurographics Symp. on Rendering}, 2005.

\bibitem[DHvOS00]{Deussen:2000:FPA}
Oliver Deussen, Stefan Hiller, Cornelius van Overveld, and Thomas Strothotte.
\newblock Floating points: A method for computing stipple drawings.
\newblock {\em Computer Graphics Forum}, 19(3), 2000.

\bibitem[DOM{\etalchar{+}}01]{Durand:2001:DSA}
Fr{\'{e}}do Durand, Victor Ostromoukhov, Mathieu Miller, Francois Duranleau,
  and Julie Dorsey.
\newblock Decoupling strokes and high-level attributes for interactive
  traditional drawing.
\newblock In {\em Rendering Techniques 2001: 12th Eurographics Workshop on
  Rendering}, pages 71--82, 2001.

\bibitem[EL99]{Efros:1999}
Alexei~A. Efros and Thomas~K. Leung.
\newblock Texture synthesis by non-parametric sampling.
\newblock In {\em IEEE International Conference on Computer Vision}, pages
  1033--1038, 1999.

\bibitem[ESM{\etalchar{+}}91]{bb3357}
A.~Etemadi, J.P. Schmidt, G.~Matas, J.~Illingworth, and J.V. Kittler.
\newblock Low-level grouping of straight line segments.
\newblock In {\em British Machine Vision Conf.}, pages 119--126, 1991.

\bibitem[FTP03]{Freeman:2003:style}
William~T. Freeman, Joshua~B. Tenenbaum, and Egon~C. Pasztor.
\newblock Learning style translation for the lines of a drawing.
\newblock {\em ACM Trans. Graph.}, 22(1):33--46, 2003.

\bibitem[Her98]{Hertzmann:1998:PRW}
Aaron Hertzmann.
\newblock Painterly rendering with curved brush strokes of multiple sizes.
\newblock In {\em Proceedings of SIGGRAPH 98}, pages 453--460, 1998.

\bibitem[HOCS02]{Hertzmann:2002:CA}
Aaron Hertzmann, Nuria Oliver, Brian Curless, and Steven~M. Seitz.
\newblock Curve analogies.
\newblock In {\em Rendering Techniques 2002: 13th Eurographics Workshop on
  Rendering}, pages 233--246, June 2002.

\bibitem[JEGPO02]{Jodoin:2002:HBE}
Pierre-Marc Jodoin, Emric Epstein, Martin Granger-Pich{\'{e}}, and Victor
  Ostromoukhov.
\newblock Hatching by example: a statistical approach.
\newblock In {\em Proceedings of NPAR 2002}, pages 29--36, 2002.

\bibitem[KMM{\etalchar{+}}02]{Kalnins:2002:WND}
Robert~D. Kalnins, Lee Markosian, Barbara~J. Meier, Michael~A. Kowalski,
  Joseph~C. Lee, Philip~L. Davidson, Matthew Webb, John~F. Hughes, and Adam
  Finkelstein.
\newblock {WYSIWYG NPR}: Drawing strokes directly on 3{D} models.
\newblock {\em ACM Trans. on Graphics}, 21(3):755--762, July 2002.

\bibitem[Llo82]{lloyd1}
S.~P. Lloyd.
\newblock Least squares quantization in pcm.
\newblock {\em IEEE Trans. on Information Theory}, 28(2):129--137, 1982.

\bibitem[Ost99]{Ostromoukhov:1999:DFE}
Victor Ostromoukhov.
\newblock Digital facial engraving.
\newblock In {\em Proceedings of SIGGRAPH 99}, pages 417--424, 1999.

\bibitem[SABS94]{Salisbury:1994:IPI}
Michael~P. Salisbury, Sean~E. Anderson, Ronen Barzel, and David~H. Salesin.
\newblock Interactive pen-and-ink illustration.
\newblock In {\em Proceedings of SIGGRAPH 94}, pages 101--108, 1994.

\bibitem[Tur01]{Turk:2001:TSO}
Greg Turk.
\newblock Texture synthesis on surfaces.
\newblock In {\em Proceedings of ACM SIGGRAPH 2001}, pages 347--354, 2001.

\bibitem[WL01]{Wei:2001:TSO}
Li-Yi Wei and Marc Levoy.
\newblock Texture synthesis over arbitrary manifold surfaces.
\newblock In {\em Proceedings of ACM SIGGRAPH 2001}, pages 355--360, 2001.

\bibitem[WS94]{Winkenbach:1994:CPI}
Georges Winkenbach and David~H. Salesin.
\newblock Computer-generated pen-and-ink illustration.
\newblock In {\em Proceedings of SIGGRAPH 94}, pages 91--100, 1994.

\end{thebibliography}


\end{document}